# A reduced-temperature process for preparing atomically clean Si(100) and SiGe(100) surfaces with vapor HF

Luis Fabián Peña,[†,‡] Evan M. Anderson,[*,†] John P. Mudrick,[†] Samantha G. Rosenberg,[†,¶] David A. Scrymgeour,[†] Ezra Bussmann,[†] and Shashank Misra[†]

[†]*Sandia National Laboratories, Albuquerque NM, 87185 USA*
[‡]*Current affiliation: Centre of Excellence for Quantum Computation and Communication Technology, School of Physics, University of New South Wales, Sydney, NSW 2052, Australia*
[¶]*Current affiliation: Lockheed Martin Space, Palo Alto, CA, USA*

E-mail: emander@sandia.gov

## Abstract

Silicon processing techniques such as atomic precision advanced manufacturing (APAM) and epitaxial growth require surface preparations that activate oxide desorption (typically >1000 °C) and promote surface reconstruction toward atomically-clean, flat, and ordered Si(100)-2×1. We compare aqueous and vapor phase cleaning of Si and Si/SiGe surfaces to prepare APAM-ready and epitaxy-ready surfaces at lower temperatures. Angle resolved X-ray photoelectron spectroscopy (ARXPS) and Fourier transform infrared (FTIR) spectroscopy indicate that vapor hydrogen fluoride (VHF) cleans dramatically reduce carbon surface contamination and allow the chemically prepared surface to reconstruct at lower temperatures, 600 °C for Si and 580 °C for a Si/Si$_{0.7}$Ge$_{0.3}$ heterostructures, into an ordered atomic terrace structure indicated by scanning tunneling microscopy (STM). After thermal treatment and vacuum hydrogen termination, we demonstrate STM hydrogen desorption lithography (HDL) on VHF-treated Si samples, creating reactive zones that enable area-selective chemistry using a thermal budget similar to CMOS process flows. We anticipate these results will establish new pathways to integrate APAM with Si foundry processing.

## Introduction

Techniques like area-selective deposition have gained increasing interest due to the inherent self-aligned fabrication scheme via control of the surface chemistry.[1,2] One promising technique is atomic precision advanced manufacturing (APAM), which enables exploration of unique device physics on atomically clean and reconstructed Si terraces.[3] APAM can produce metallic regions of strongly doped silicon by exposing scanning tunneling microscope (STM)patterned regions of hydrogen-terminated silicon to dopant precursors.[4–6] This ultra-high doping modifies the band structure[7] and enables higher current density transport comparable to that of copper[8] enabled by multiple bands available for transport.[9] A challenge for integrating APAM into Si MOS, SiGe, and other platforms is the thermal requirement for preparing atomically-clean Si surfaces.[10,11]

Typical APAM surface preparations rely on annealing at approximately 1000 °C or more in ultra-high vacuum environments. Substantial prior effort focused on finding lower-T surface preparation techniques.[12,13] For example, hydrofluoric acid (HF) processing combined with ion sputtering in vacuum yielded an APAM-ready surface for $T \geq 600$ °C.[12] In many processes, aqueous HF etching has been the preferred treatment to remove surface oxides and induce hydrogen termination prior to subsequent processing such as microelectronics fabrication, epitaxy, and surface functionalization.[14–19] Similarly, gas-phase HF mixtures with aliphatic alcohols or acetic acid have also been explored for surface oxide removal and selectively etching oxides for complex structures.[20–25]

Despite its ubiquitous use, the interaction of Si with vapor HF (VHF) has not yet been extensively characterized, or optimized for applications like APAM. We will bridge this gap by characterizing the interaction of Si with VHF using infrared spectroscopy and X-ray photoelectron spectroscopy to determine its suitability for applications like APAM. The advantages of vapor phase etching over conventional aqueous routes include the elimination of mobile ions, organic, inorganic, and metallic impurities typically found in solution.[26] For example, anhydrous HF/water vapor has been demonstrated to achieve stiction-free microelectromechanical system devices, otherwise, not possible via aqueous routes.[24,27–29] VHF etching has the potential to be more efficient and less aggressive, resulting in smoother surfaces,[30] and does not require a deionized (DI) water rinse, which has been reported as a source for C contamination.[26,31–34] For instance, Wong et al. investigated the effectiveness of vapor HCl/HF mixtures to clean the surface before oxide growth and reported an improvement in the endurance of the grown oxide in capacitors compared to conventional wet cleans.[35] Unfortunately, the cleaned surface was not studied directly, which makes it impossible to understand how the vapor treatments affect the surface termination. A systematic investigation is needed to close the gap in knowledge and make the connection to APAM processing.

Here, we investigate the surface chemical effects of VHF on Si(100) substrates using Fourier transform infrared spectroscopy (FTIR) and angle resolved X-ray photoelectron spectroscopy (ARXPS) to achieve a suitable surface termination. We found that using VHF produces cleaner surfaces compared to the traditional aqueous process. VHF treated Si(100) resulted in passivating Si-H$_2$ and Si-F species that suppress oxide formation. After annealing at 580 °C to foster large terrace reconstruction, the surface underwent structural changes resulting in an epi-ready surface but lost its chemical hydrogen termination. The VHF method was also tested on a more delicate surface, Si quantum wells on Si$_{0.7}$Ge$_{0.3}$ virtual substrate, and produced similar results. We demonstrate suitably clean Si and Si/SiGe surfaces for APAM that are compatible with Si CMOS processing, and demonstrate STM hydrogen desorption lithography (HDL) that enables area-selective bottom-up fabrication chemistries. The trend toward integrated systems, e.g., cluster tools, makes vapor phase acid exposures appealing to prepare and maintain clean surfaces critical for Si fabrication schemes.[3,23,36–38]

## Experimental

In this study, 6 inch, double sided polished, FZ Si(100) wafers (p-type) were diced to 25×25 mm$^2$ for FTIR, and 5×9 mm$^2$ chips for STM and ARXPS measurements. Sample preparation involved subsequent sonication in acetone, methanol, and isopropanol for 10 min each, then drying with

N₂ gas. Next, samples were treated with UV ozone for 20 min. For wet vs. vapor treatment comparison, wet etching consisted of reproducing the optimized hydrogen termination procedure for Si(100), hereafter referred to as Clean 1.[39] The hydrogen termination of Clean 1 was compared to the well-known and optimized hydrogen termination for Si(111) used by Higashi et al, referred to as Clean 2.[40] VHF experiments were carried out by first treating substrates with a Piranha solution (3:1 $H_2SO_4$:$H_2O_2$ at 85 °C) for 10 min, a dip in DI water and drying with N₂ gas. Samples were suspended face down for 15 to 120 s above a beaker filled with aqueous 49% HF at room temperature inside a fume hood. For FTIR measurements, both sides of the Si(100) sample were treated with VHF. Si/SiGe heterostructures on $Si_{0.7}Ge_{0.3}$ virtual substrate, purchased from Lawrence Semiconductor Research Labs, underwent similar VHF cleaning procedures as blank silicon and were only characterized with STM.

### Angle Resolved X-ray Photoelectron Spectroscopy (ARXPS)

Prior to ARXPS measurements, samples were exposed to ambient conditions for ~30 min while mounting onto the sample holder and transferring into the N₂ filled load lock. The ARXPS spectra were obtained using a Kratos RSF spectrometer with a monochromatic Al K$\alpha$ source operating at a power of 200 mW, a grazing incidence angle of 60°, and a pass energy of 100 eV and 20 eV for survey and high-resolution scans, respectively. A base pressure of ~$10^{-10}$ Torr was maintained throughout ARXPS analysis. The associated CASA software was used for peak fitting and analysis. Peaks were aligned to the carbon 1s peak at 285 eV and a Shirley background subtraction was used to fit the peaks.

### Fourier Transform Infrared Spectroscopy (FTIR)

Prior to FTIR measurements, samples were exposed to ambient conditions 30 min while mounting into the sample holder and loading into N₂-purged instrument. Spectra were collected with a Thermo Nicolet 6700 FTIR spectrometer equipped with a KBr beamsplitter, Globar IR source, and deuterated triglycine sulfate detector. Absorbance spectra were acquired over 400–4000 cm$^{-1}$ with a resolution of 4 cm$^{-1}$ using a single-pass transmission geometry. The incidence angle was kept close to the Brewster angle (74°) to enhance transmission, minimize interference, and enhance sensitivity to vibrational absorptions polarized parallel and perpendicular to the surface. Three loops of 500 scans each were collected for each sample and averaged for data processing.

### Scanning Tunneling Microscopy (STM)

Chemically prepared hydrogen terminated Si(100) samples were loaded within 10 min into a VT-STM (ScientaOmicron Nanotechnology) system with base pressure of 2 × $10^{-10}$ Torr. Prior to sample preparation, sample holder (S218201-S from Omicron) and manipulator stage were degassed overnight at 750 °C via radiative heating to minimize the level of outgassing during sample processing. Samples were annealed to 350 °C via radiative heating (ramping rate of 1 °C/s) to degas for 10 min prior to any treatment. Immediately after, samples were annealed to 600 °C via Joule heating (at 1 °C/s) for 30 min and then cooled to room temperature (at 5 °C/s). Temperatures were recorded with a band edge thermometer (kSA BandiT) with an accuracy of ±

2 °C. Samples were annealed at 350 °C while exposed to 2700 L of atomic hydrogen using a tungsten filament to crack $H_2$. The hydrogen pressure was controlled using a leak valve and measured through an ion gauge; exposures were calculated in terms of Langmuirs [1 Langmuir (L)= $1\times10^{-6}$ Torr during 1 s]. Constant-current STM ($I_{sp}$ = 200 pA) was operated with a sample bias between -1.3 and -2.0 V for atomic-scale surface imaging.

## Results and Discussion

To evaluate the chemical suitability of VHF and Clean 1 for preparing clean Si(100) surfaces for APAM, samples were first analyzed using FTIR and ARXPS after each clean. FTIR is one of the few characterization techniques sensitive enough and capable of detecting H non-invasively. Analysis of the intensity and linewidth of Si-H vibrational modes in the FTIR spectra was used to characterize the homogeneity of surface chemical species. Fig. 1a shows the IR absorbance spectra for four samples subjected to different VHF exposure times to establish the timescale for full surface oxide removal. We found that a minimum VHF exposure time of 60 seconds was needed to completely etch the surface chemical oxide, as demonstrated by a lack of change in the two dominant vibrations at 2141 and 2110 cm$^{-1}$, which correspond to Si(100) surface dihydride stretching modes.[39,41–44] Figure 1b compares the IR spectra of VHF and Clean 1, and the corresponding vibrational modes are identified in Table 1. The appearance of additional vibrational modes in the Clean 1 spectrum implies that Clean 1 exposes various facets on the Si(100) surface, and produces additional bonding configurations, that are observed in IR spectra as Si-H absorbance bands. In particular, the mode at 2128 cm$^{-1}$ in the Clean 1 spectrum has been previously reported and attributed to exposed Si(111) microfacets due to $H_2$ gas evolution during etching.[45] For both VHF and Clean 1, the FTIR data exhibit inhomogeneous broadening of the Si-H modes, indicating atomically rough surfaces with multiple Si-H$_x$ modes.

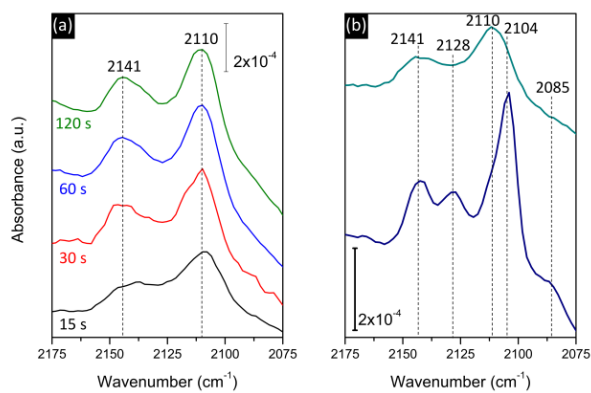

Figure 1: Differential infrared absorbance spectra showing (a) four Si(100) samples subjected to different VHF exposure times and (b) spectra for optimized VHF exposure (top) and Clean 1 (bottom) treated Si(100). It is evident that VHF exposures ≥ 60 s are necessary to etch the oxide and saturate the surface.

Table 1: Infrared vibrational mode assignment for observed Si(-H)x on VHF and Clean 1 treated Si(100) samples from calculated and experimental values reported in literature

| Wavenumber (cm$^{-1}$) | Facet | Type | VHF (*Si–H stretch*) | Clean 1 (*Si–H stretch*) | Lit. ref. |
|---|---|---|---|---|---|
| 2085 | (100) | Dihydride | n/o | Strained, down Si-H | 39,41–43 |
| 2104 | (100) | Dihydride | Unstrained, sym. | Unstrained, sym. | 39,41,43 |
| 2110 | (100) | Dihydride | Unstrained, antisym. | Unstrained, antisym. | 39,41–44 |
| 2128 | (111) | Monohydride | n/o | Strained | 39,41,43 |
| 2141 | (100) | Dihydride | Strained, up Si-H | Strained, up Si-H | 39,41,43 |

n/o = not observed; sym. = symmetric; antisym. = antisymmetric

To correlate the linewidth of the Si-H modes observed in IR spectroscopy to surface structure obtained from aqueous or vapor phase etch, we investigated the effect of VHF on Si(111). It is well known that the anisotropic etch of aqueous NH$_4$F on Si(111) produces large terraces ideally terminated with hydrogen.[40,46] The infrared spectrum of the monohydride termination, Si-H stretch at 2083 cm$^{-1}$, is a single sharp band with a linewidth of 0.9 cm$^{-1}$ at room temperature (0.1 cm$^{-1}$ at T = 130 K).[47] Interestingly, after performing the same VHF treatment on Si(111), we observed the same inhomogeneous broadening of Si-H modes as on Si(100), see Fig. 2. From these results, we correlate the intensity and linewidth to the homogeneity of Si-H and surface perfection, i.e., less intense and broadened vibrational modes indicate an atomically rough surface with multiple Si-H$_x$ configurations.

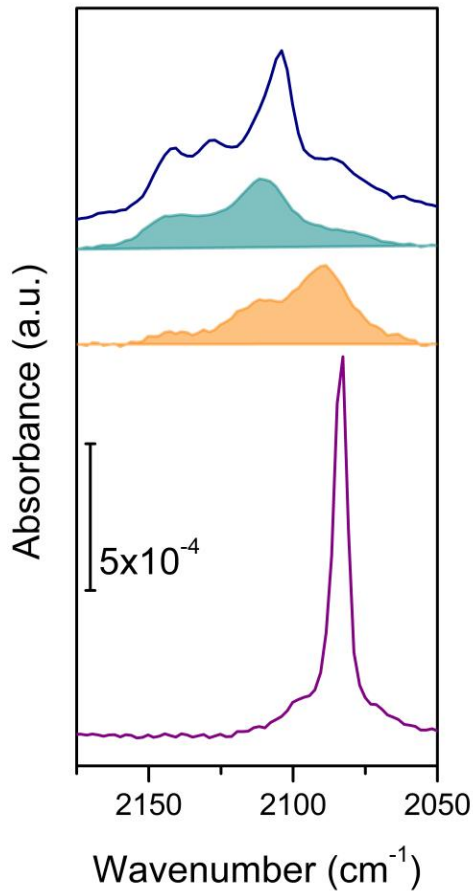

Figure 2: Differential infrared absorbance spectra for four samples (2 Si(100) and 2 Si(111)) subjected to either Clean 1, Clean 2 (lines) or VHF (shaded lines). Starting from top of figure: Clean 1 treated Si(100), VHF treated Si(100), VHF treated Si(111), and Clean 2 treated Si(111). The degree of atomic roughness can be inferred based on the intensity and linewidth of the H-Si vibrational stretching mode, i.e., a broad vibrational mode indicates an atomically rough surface with a high density of Si islands terminated with hydrogen, as observed for both VHF treated silicon samples. Whereas, a narrow and sharp mode indicates larger atomically smooth terraces with HSi modes that are coupled together and closely packed, as observed for the well-known ideally terminated H-Si(111), purple curve. The vibrational mode observed from Clean 2 treated Si(111) belongs to the monohydride terminated surface and is centered at 2083 cm$^{-1}$. Clean 2 reproduces the optimized Si(111) cleaning procedure used by Higashi et al.[40]

As a complementary technique to FTIR, ARXPS measurements at grazing incidence were carried out on Si(100) samples after the VHF and Clean 1 processes for an elemental profile of surface contaminants. Figure 3 shows a comparison of the presence of F, O, and C after each sample was cleaned. An elemental survey of each surface is provided in Supplementary Information (SI) Figure 1. ARXPS results indicate that only Clean 1 contained fluorocarbon contamination, evident by the strong F and C signals. For VHF, F contamination was in the form of fluorinated silicon. For both surface cleans, the presence of oxygen was attributed to carbonate and moisture adsorption during sample transport, rather than unetched $SiO_2$. Thus, the VHF process produced a cleaner Si(100) surface than Clean 1.

The presence of Si-F on VHF treated Si(100) samples stems from the exhaustion of surface moisture created during the etch after the oxide has been consumed. It is well known that during the DI aqueous rinse, $H_2O$ attacks F-Si(100) to form HO-Si(100) and HF byproduct in a matter of seconds; without water, there is nothing to react with Si-F.[14] Yet, if the DI water source is not ultra-clean, then the immersion in an aqueous etching solution or rinsing process could be a source of C contamination that produces fluorocarbon species, as demonstrated in Fig. 3. We acknowledge that there may be some contamination from exposure to ambient air that contributes to the formation of fluorocarbon. However, we do not believe ambient air to be the main cause of C contamination since we observe differing levels in comparing the Clean 1 and VHF processes. Based on this, it is more probable that immersion in an aqueous solution or the use of DI water for rinsing after Clean 1 is the primary source. The low contamination of Si-F is not concerning as it may be possible to desorb as volatile $H_2SiF_6$, or other analog, at T ≥ 300 °C,[48] and it can actually help prevent oxide growth during sample transfer.[42]

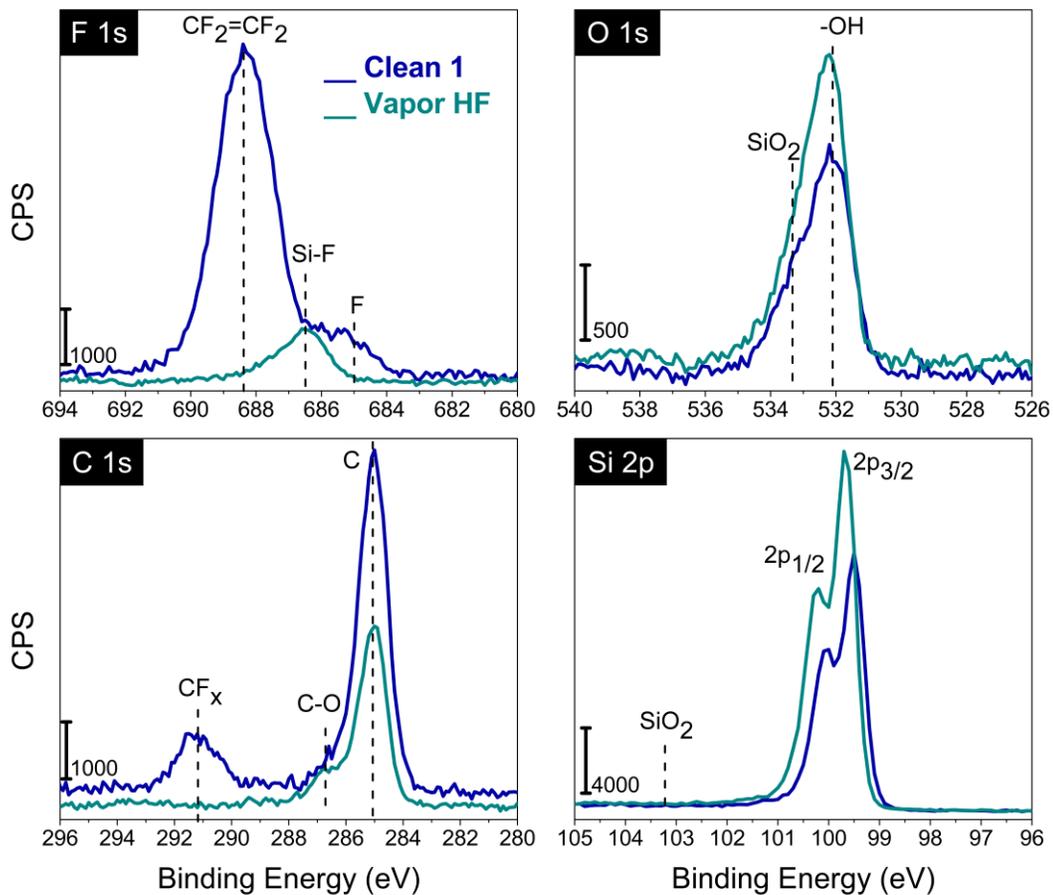

Figure 3: ARXPS at grazing incidence for F, O, C, and Si core levels for VHF and Clean 1 treated Si(100) samples showing less C and F contamination with VHF than Clean 1. O levels are attributed to adsorbates during sample transport rather than incomplete $SiO_2$ removal.

The atomic surface structure of samples after the VHF process was investigated using STM to evaluate the completeness of oxide removal and the extent of residual contamination after annealing samples in vacuum. Fig. 4 shows the reconstructed atomic surface structure of two Si(100) samples, Sample 1 and Sample 2, after VHF treatment and annealing at 580 °C for 30 min (removing any H), confirming the FTIR result that the oxide had been completely removed. Sample 1 was flashed to > 1100 °C to examine the extent of C contamination, which manifests as surface carbides that form tall mounds and pin Si atomic steps, [13] see SI Fig. 2. In Fig. 4(b), the lack of evidence of surface carbides in the image indicates an atomically clean and relaxed Si surface. Fig. 4c displays the surface of Sample 2 after annealing to 580 °C and H-terminating *in-situ*, while Fig. 4d shows the surface after STM lithography. The bright contrast in Fig. 4d, chemical contrast, is the area where H has been desorbed resulting in Si dangling bonds, which are the regions enclosed by the yellow markers where selective reactions can take place. Thus, after annealing, the VHF process is suitable for subsequent hydrogen termination, H desorption lithography, and the full APAM process.

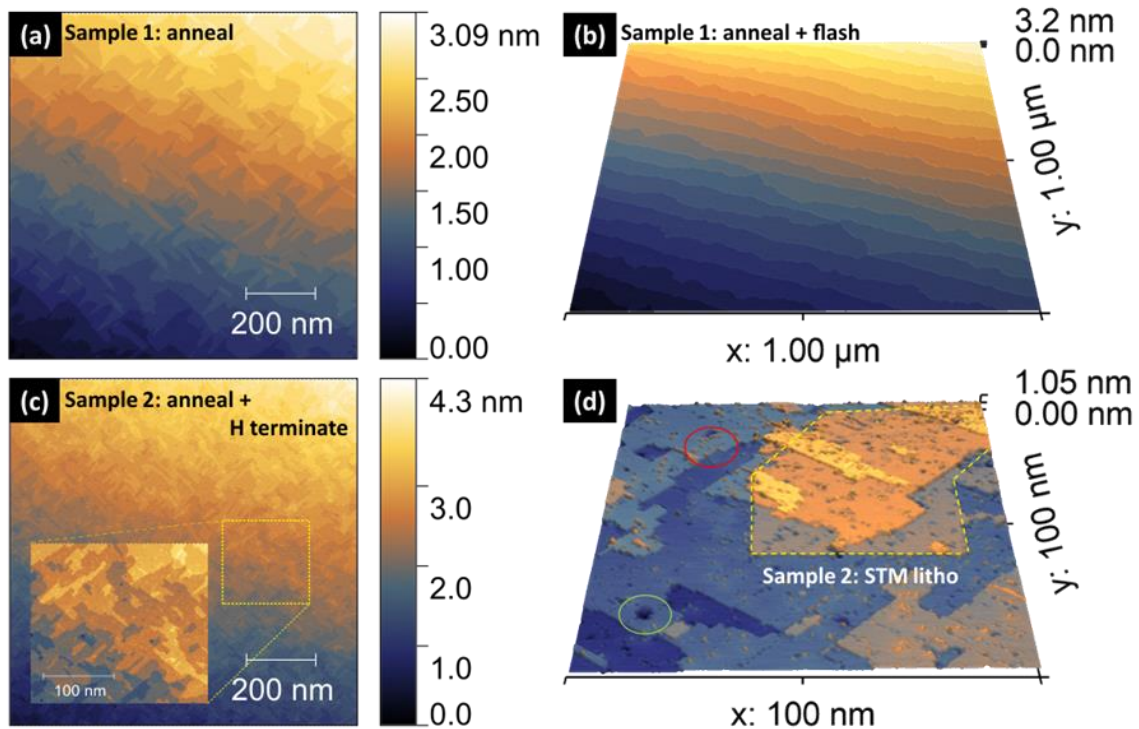

Figure 4: STM images of two VHF treated Si(100) samples, Sample 1 and Sample 2, after (a) annealing Sample 1 to 600 °C and (b) flashing to >1100 °C. (c) Annealing Sample 2 to 600 °C and H-terminating, and (d) post hydrogen desorption lithography on monohydride terminated Si(100), dashed yellow lines surround STM lithography. (d) Captures certain types of surface defects, such as adatoms which appear as pits after reconstruction (green circle), and unavoidable stray dangling bonds (red circle).

The VHF treated surface was rough at an atomic level that it was challenging to detect any chemical contrast even after attempting lithography, which only creates a height difference of ~0.1 nm for patterned vs. unpatterned areas. To clean and reconstruct the Si surface, an anneal process was needed. This involved heating the surface to 250 °C to desorb any physisorbed contamination and transform dihydride to monohydride. As the temperature exceeds 420 °C, monohydrides also desorb and to enable STM lithography the surface needs to be hydrogen-terminated *in-situ* with a hydrogen cracker.[49] Long-range (2×1) reconstruction of Si(100) was not apparent until the annealing temperature reached ≥ 550 °C. After reconstructing and hydrogen terminating, STM lithography successfully exposed selective areas for chemical reactions. The annealing temperature can be lowered to the desorption temperature of monohydride-terminated Si(100)-2×1. Below this range, hydrogen hinders Si mobility and restricts the formation of large terraces. At these lower temperatures, small islands may form, making it difficult to distinguish lithographic patterns. In such cases, $dI/dV$ mapping may be necessary.

Additionally, we investigated the potential of VHF to clean Si quantum wells on Si/Si$_{0.7}$Ge$_{0.3}$ heterostructures, which have a lower thermal budget requirement than Si, ≪ 700 °C, to minimize surface roughening.[50,51] Si/SiGe samples had been exposed to ambient conditions for over a year, but with VHF treatment, we were able to clean and reconstruct the surface using the same process, Fig. 5. These instances demonstrate the effectiveness of VHF, particularly when the acid etch is performed without submerging the sample in aqueous solutions.

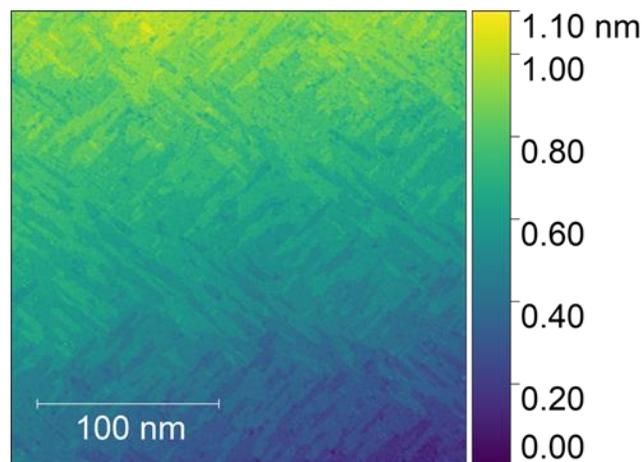

Figure 5: STM image of Si/Si$_{0.7}$Ge$_{0.3}$ heterostructure surface after VHF treatment and annealing to 580 °C showing 250 nm × 250 nm scan displaying the atomic step distribution. Refer to SI Fig. 3 for large area scan of Si/SiGe surface.

As discussed above, the results presented here demonstrate a viable pathway for preparing epi-ready Si surfaces after performing a mild anneal that meets the thermal requirements for back-end-of-line processing. Additionally, these surfaces can be H-terminated *in situ*, enabling the possibility of hydrogen desorption lithography. The advantage of VHF treatments is that this process can be scaled up to wafer dimensions, surpassing other techniques like surface sputtering, which can introduce impurities, roughen surfaces, and are challenging to scale up. This process can replace aqueous HF treatments and be integrated into a dedicated chamber as part of a larger cluster tool, without the need to expose the surfaces to ambient conditions prior to subsequent processes such as epitaxy/deposition or chemical surface functionalization.

## Conclusions

We investigated low temperature chemical processes for removing the oxide from Si(100) to provide surfaces suitable for APAM device processing. We found that using VHF produces cleaner surfaces at reduced temperatures as low as 580 °C as compared to the traditional aqueous process. We used two complementary techniques, FTIR and ARXPS, to understand the surface termination and chemical composition. VHF treated Si(100) resulted in passivating $SiH_2$ and Si-F species that suppress oxide formation and can be desorbed at T ≥ 300 °C.[48] After annealing at 600 °C, the surface underwent structural changes but lost its chemical hydrogen termination. By hydrogen terminating *in-situ* and employing STM lithography, we were able to selectively control reactions in specific areas. The trend toward integrated systems, e.g., cluster tools, makes vapor phase acid exposures appealing to prepare and maintain clean surfaces critical for device fabrication. This allowed us to advance APAM without the need for high-temperature preparation, reduced to 600 °C, greatly improving integration possibilities between CMOS Front-end-of-line (FEOL) and the CMOS Back-end-of-line (BEOL). Monolithic integration of APAM components into CMOS circuits opens the door for devices with enhanced functionality. Further, this HF vapor clean was suitable for reconstructing the surface of $Si/Si_{0.7}Ge_{0.3}$ heterostructures at low temperature, making it applicable to optimizing the growth and fabrication of spin qubits.

## Acknowledgement

The authors thank Clare Davis-Wheeler Chin, Scott Schmucker, and Jeff Ivie for helpful discussions. This work was supported by the Laboratory Directed Research and Development Program at Sandia National Laboratories under project 213017 and was performed, in part, at the Center for Integrated Nanotechnologies, a U.S. DOE, Office of Basic Energy Sciences user facility. This article has been authored by an employee of National Technology and Engineering Solutions of Sandia, LLC under Contract No. DE-NA0003525 with the U.S. Department of Energy (DOE). The employee owns all rights, title and interest in and to the article and is solely responsible for its contents. The U.S. Government retains, and the publisher, by accepting the article for publication, acknowledges that the U.S. Government retains a nonexclusive, paid-up, irrevocable, worldwide license to publish or reproduce the published form of this manuscript or allows others to do so, for U.S. Government purposes. The Department of Energy will provide public access to these results of federally sponsored research in accordance with the DOE Public Access Plan https://www.energy.gov/doe-publicaccess-plan. This paper describes objective technical results and analysis. Any subjective views or opinions that might be expressed in the paper do not necessarily represent the views of the U.S. Department of Energy or the U.S. Government.

# Supplementary Information

ARXPS survey of elemental states for Si(100) samples treated with Clean 1 and VHF. STM of aqueous HF treated Si(100) surface showing carbide formation and Si/Si$_{0.7}$Ge$_{0.3}$ heterostructure surface after VHF treatment and annealing to 580 °C.

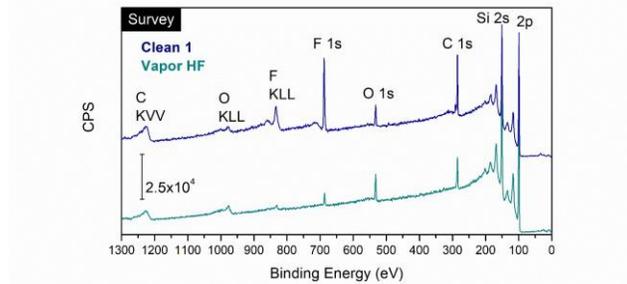

SI Fig. 1. ARXPS survey of elemental states for Si(100) samples treated with Clean 1 (top) and vapor HF (bottom).

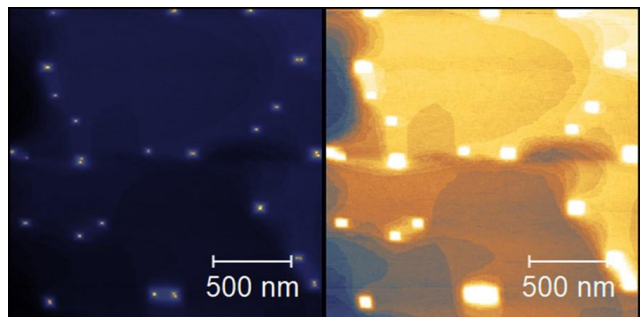

SI Fig. 2. STM image of aqueous HF treated Si(100) showcasing (left) carbide formation and (right) step pinning after flashing at T > 800 °C. Right image was processed to highlight Si terrace features.

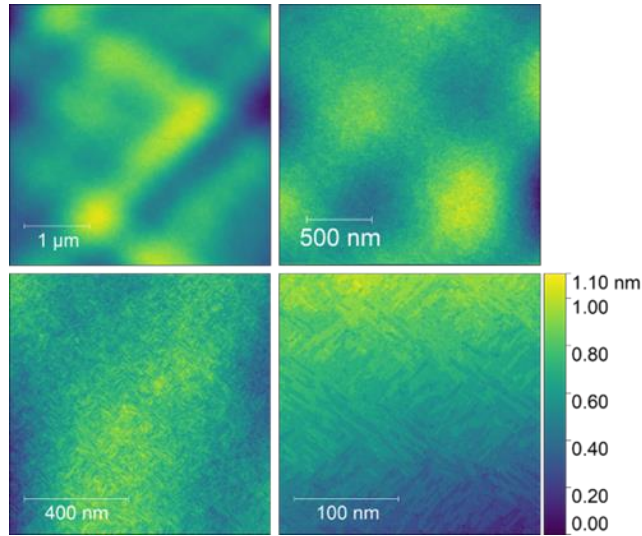

SI Fig. 3. STM image of Si/Si$_{0.7}$Ge$_{0.3}$ heterostructure surface after vapor HF treatment and annealing to 580 °C showing (top-left) 4 μm × 4μm scan highlighting part of the cross-hatch pattern, and (bottom-right) 250 nm × 250 nm scan displaying the atomic step distribution.

## TOC Graphic

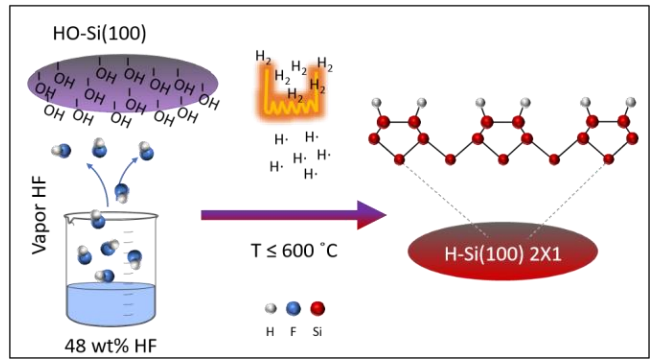